\documentclass{aa}  
\usepackage{graphicx}
\usepackage{txfonts}
\usepackage{natbib,twoopt}
\usepackage{soul}
\bibpunct{(}{)}{;}{a}{}{,}             %% natbib format for A&A and ApJ
\usepackage{hyperref}
\hypersetup{
    colorlinks=true,
    linkcolor=blue,
    citecolor = blue,
    filecolor=magenta,      
    urlcolor=blue}

\begin{document} 

\authorrunning{Alvarez Garay et al.}
\titlerunning{X-shooter spectroscopy of Liller~1 giant stars}

\title{X-shooter spectroscopy of Liller~1 giant stars 
\thanks{Based on observations collected at the Very Large Telescope (VLT) of the European Southern Observatory at Cerro Paranal (Chile) under program 089.D-0306 (PI:Ferraro).}}

\author{D. A. Alvarez Garay\inst{1}\fnmsep\inst{2},
    C. Fanelli\inst{2},
    L. Origlia\inst{2},
    C. Pallanca\inst{1}\fnmsep\inst{2},
    A. Mucciarelli\inst{1}\fnmsep\inst{2},
    L. Chiappino\inst{1}\fnmsep\inst{2},
    C. Crociati\inst{1}\fnmsep\inst{2}, 
    B. Lanzoni\inst{1}\fnmsep\inst{2},
    F. R. Ferraro\inst{1}\fnmsep\inst{2},
    R. M. Rich\inst{3},
    and E. Dalessandro\inst{2}}

   \institute{Dipartimento di Fisica e Astronomia, Università degli Studi di Bologna, Via Gobetti 93/2, I-40129 Bologna, Italy\\
\email{deimer.alvarezgaray2@unibo.it}
        \and
       INAF, Osservatorio di Astrofisica e Scienza dello Spazio di Bologna, Via Gobetti 93/3, I-40129 Bologna, Italy
         \and
        Department of Physics and Astronomy, UCLA, 430 Portola Plaza, Box 951547, Los Angeles, CA 90095-1547, USA}

\abstract
{We present the first comprehensive chemical study of a representative sample of 27 luminous red giant branch (RGB) stars belonging to Liller~1, a complex stellar system in the Galactic bulge. This study is based on medium-resolution near-infrared spectra acquired with X-shooter at the Very Large Telescope. 
We found a subpopulation counting 22 stars with subsolar metallicity ($<$[Fe/H]$>=-0.31\pm0.02$ and 1$\sigma$ dispersion of 0.08 dex) and with enhanced [$\alpha$/Fe], [Al/Fe], and [K/Fe] that likely formed early and quickly from gas that was mainly enriched by type~II supernovae, and a metal-rich population counting 5 stars with supersolar metallicity ($<$[Fe/H]$>$=+0.22$\pm$0.03 and 1$\sigma$ dispersion of 0.06 dex) and roughly solar-scaled [$\alpha$/Fe], [Al/Fe], and [K/Fe] that formed at later epochs from gas that was also enriched by type~Ia supernovae. Moreover, both subpopulations show enhanced [Na/Fe], as in the bulge field, about solar-scaled [V/Fe], and depletion of [C/Fe] and $^{12}$C/$^{13}$C with respect to the solar values. This indicates that mixing and extra-mixing processes during the RGB evolution also occur at very high metallicities.
Notably, no evidence of a Na-O anticorrelation, which is considered the fingerprint of genuine globular clusters, has been found. This challenges any formation scenarios that invoke the accretion of a molecular cloud or an additional stellar system onto a genuine globular cluster.
The results of this study underline the strong chemical similarity between Liller 1 and Terzan 5 and support the hypothesis that these complex stellar systems might be fossil fragments of the epoch of Galactic bulge formation.}


\maketitle
%
%-------------------------------------------------------------------
\section{Introduction} \label{intro}

\begin{figure*}
    \centering
    \includegraphics[width=1\linewidth]{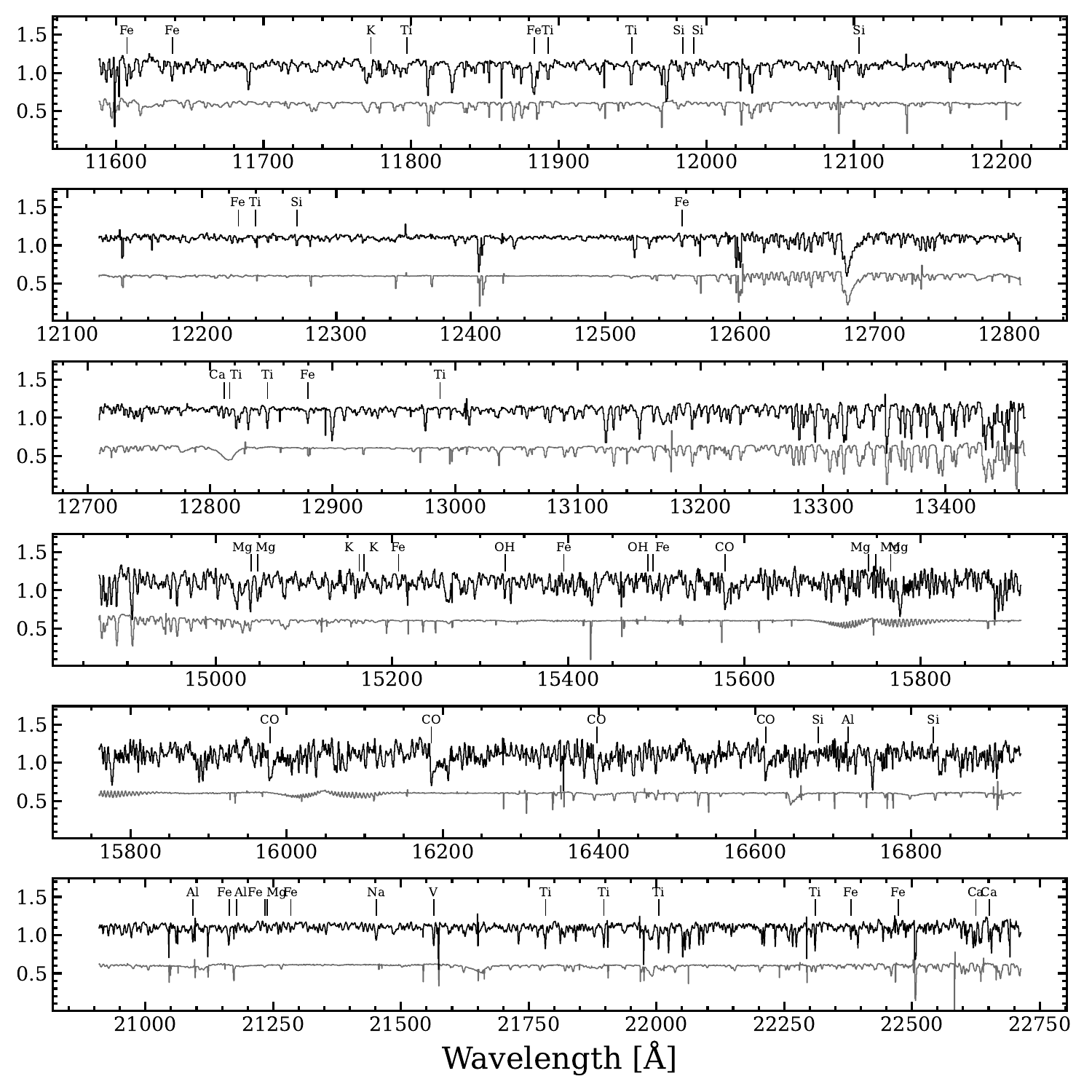}
    \label{spectral_lines}
    \caption{ Portions of the X-shooter spectrum for a giant star with a temperature of 3400 K across the J, H, and K bands (featuring X-shooter orders 22, 21, 20, 17, 16, and 12 from top to bottom). Some lines of interest for the chemical analysis are also marked. In each panel, the upper spectrum in black is the stellar normalized spectrum corrected for radial velocity, and the lower gray spectrum shows the telluric absorption.}
\end{figure*}

Liller~1 is a massive stellar system in the Galactic bulge located at 0.8 kpc from the Galactic center. It has an  average E(B-V) color excess of $\approx$4.5 mag and a differential reddening $\delta$E(B-V) that ranges between -0.57 and +0.37 mag \citep{Pallanca_21}.
Because of its high absolute and differential reddening, Liller~1 has remained relatively unexplored until very recently.

A high-quality, deep color-magnitude diagram (CMD) from a combination of I- and K-band images acquired with the Hubble Space Telescope (HST) and the Gemini South Adaptive Optics Imager (GSAOI) camera at the Gemini South telescope, respectively, has revealed two subpopulations within Liller~1: A main 12 Gyr old and likely subsolar subpopulation, and a younger (only 1-3 Gyr old) and likely supersolar population \citep{ferraro_21}.
This high-quality CMD has been also used as reference to model the Liller~1 star formation history (SFH; \citealt{dalessandro_22}).
The best-fit solution reveals that star formation was active in Liller~1 for nearly its entire existence, with three distinct episodes. The main episode started approximately 12-13 Gyr ago, with a tail extending for about 3 Gyr, and it created approximately 70\% of the current total mass of the system.
The second episode occurred between 6 and 9 Gyr ago, contributing an additional approximately 15\% of the system mass.
The most recent star formation event started around 3 Gyr ago and ceased roughly $1$ Gyr ago, when a quiescent phase began. This young subpopulation constitutes $\approx10\%$ at least of the total mass of Liller~1.

The only high-resolution IR spectroscopic study of Liller~1 to date has been performed with Near InfraRed Spectrograph (NIRSpec) at Keck, providing iron and $\alpha$-element abundances for two giant stars that belong to the metal-poor ([Fe/H]=-0.3 dex) and $\alpha$-enhanced ([$\alpha$/Fe]=+0.3 dex) subpopulation \citep{origlia_02}.
The first Liller~1 metallicity distribution, recently obtained from the analysis of low-resolution spectra obtained with the Multi Unit Spectroscopic Explorer (MUSE) in the Ca triplet region \citep{Crociati_23}, shows a dominant metal-poor component with a peak at [Fe/H]$\approx-0.5$ dex and a metal-rich component with a peak at [Fe/H]$\approx+0.3$ dex. Both components have a 1$\sigma$ dispersion of about 0.2 dex.

These findings indicate that Liller~1 is another complex stellar system of the bulge, hosting multi-age and multi-iron stellar subpopulations similarly to Terzan~5 \citep[see][]{Ferraro_09, ferraro_16}. Hence, a proper spectroscopic screening of its elemental abundances is urgent in order to properly characterize its chemistry and place its subpopulations in a comprehensive evolutionary scheme, as was previously done with Terzan~5 \citep{Origlia_11, Origlia_13, origlia_19,Massari_14}.
We present a chemical study based on near-IR (NIR) X-shooter spectra of 27 stars that are likely members of Liller~1. Observations and membership from proper motions  are described in Sects.~\ref{obs} and \ref{pm}, respectively, and the spectral analysis and our results for the kinematics and chemical abundances are presented  in Sects.~\ref{analysis} and \ref{res}, respectively. In  Sect.~\ref{disc} we briefly discuss the inferred Liller~1 properties in the broader context of its possible origin and evolution. In the past years, a number of scenarios for an in situ formation or an extragalactic origin of Terzan~5 and Liller~1 have been proposed in the literature (see \citealt{brown18,alfaro19,taylor22,mckenzie18,bastian22, khoperskov18,mastrobuono19,pfeffer21,moreno22,ishchenko23}), but none of them properly accounted for all the evolutionary, chemical, and kinematic observational evidence that is available for these complex stellar systems.

%%%%%%%%%%%%%%%%%%%%%
\begin{table}
\caption{Observed stars in Liller 1.}
\label{tab:1}
\centering
\renewcommand{\arraystretch}{1.25}
\scriptsize
\setlength{\tabcolsep}{9pt}
\begin{tabular}{|c|c|c|c|c|c|}
\hline\hline 
ID & RA & Dec & J & K & distance \\
& [Deg] & [Deg] & [mag] & [mag] & [arcsec] \\ 
\hline
20   &  263.3413160  &  -33.3917580  &  12.27  &   9.15  &  34.3  \\
24   &  263.3580030  &  -33.3844380  &  12.03  &   9.26  &  25.0  \\
27   &  263.3463770  &  -33.3867450  &  11.97  &   9.22  &  20.6  \\
31   &  263.3386050  &  -33.3953320  &  12.10  &   9.29  &  46.4  \\
34   &  263.3370940  &  -33.3755720  &  12.07  &   9.30  &  68.0  \\
35   &  263.3561440  &  -33.3858760  &  12.12  &   9.28  &  17.6  \\
37   &  263.3408550  &  -33.3894200  &  12.21  &   9.36  &  34.6  \\
39   &  263.3619960  &  -33.3893320  &  12.20  &   9.26  &  28.9  \\
45   &  263.3540437  &  -33.3894655  &  12.02  &   9.48  &   5.3  \\
48   &  263.3509838  &  -33.3909300  &  12.41  &   8.98  &   6.6  \\
62   &  263.3659950  &  -33.4061700  &  12.50  &   9.75  &  72.6  \\
66   &  263.3614350  &  -33.4098700  &  12.50  &   9.78  &  78.3  \\
68   &  263.3392510  &  -33.4054300  &  12.46  &   9.78  &  69.5  \\
71   &  263.3535490  &  -33.3906021  &  12.43  &   9.87  &   5.2  \\
74   &  263.3490706  &  -33.3910950  &  12.80  &  10.01  &  11.4  \\
79   &  263.3491370  &  -33.3856470  &  12.65  &  10.02  &  17.4  \\
85   &  263.3551136  &  -33.3912034  &  12.84  &   9.95  &  10.3  \\
88   &  263.3537148  &  -33.3957827  &  12.80  &  10.01  &  22.9  \\
98   &  263.3560121  &  -33.3878820  &  12.98  &  10.41  &  12.6  \\
100  &  263.3440470  &  -33.3914300  &  13.00  &  10.38  &  26.2  \\
103  &  263.3554470  &  -33.3868030  &  13.07  &  10.33  &  13.4  \\
104  &  263.3478150  &  -33.3866650  &  13.13  &  10.49  &  17.3  \\
108  &  263.3515430  &  -33.4021070  &  13.06  &  10.31  &  45.5  \\
109  &  263.3548662  &  -33.3878013  &  13.08  &  10.60  &   9.8  \\
115  &  263.3599230  &  -33.3909680  &  13.14  &  10.36  &  22.9  \\
120  &  263.3599690  &  -33.3839110  &  13.16  &  10.44  &  30.2  \\
121  &  263.3404830  &  -33.3814510  &  13.09  &  10.47  &  46.1  \\
124  &  263.3421180  &  -33.4040030  &  13.11  &  10.49  &  60.6  \\
126  &  263.3496644  &  -33.3857118  &  13.32  &  10.71  &  16.0  \\
132  &  263.3630190  &  -33.4004900  &  13.25  &  10.65  &  50.8  \\
136  &  263.3545365  &  -33.3861169  &  13.46  &  10.78  &  13.8  \\
145  &  263.3499870  &  -33.3816380  &  13.45  &  10.83  &  29.3  \\
148  &  263.3529737  &  -33.3828074  &  13.48  &  11.15  &  24.1  \\
149  &  263.3629250  &  -33.4037590  &  13.50  &  10.85  &  60.5  \\
\hline\hline
\end{tabular}
\vspace{1pt}
\end{table}
%%%%%%%%%%%%%%%%%%%

\section{Observations} 
\label{obs}
We observed 34 bright giant stars in Liller~1 with the X-shooter spectrograph at the VLT under program 089.D-0306 (PI: F.R. Ferraro). 

Because of the huge reddening towards Liller~1, only the spectra acquired with the NIR arm of X-shooter and the 0.6 arcsec slit (providing a resolution of R$\approx$8,000) in the 1.15-2.37 $\mu$m range have a sufficiently high signal-to-noise ratio to be effectively used for a chemical analysis. 
The  X-shooter spectra were acquired by nodding on slit, with a typical throw of a few arcseconds for an optimal subtraction of the background. In particular, one AB nodding cycle was selected for a total on-source integration time ranging from 4 to 14 min, depending on the stellar brightness.

The spectra were reduced with the ESO X-shooter pipeline version 3.1.0 to obtain 2D rectified and wavelength-calibrated spectra. The 1D spectrum was extracted manually in order to optimize the location and extension on the detector of the pair of spectra corresponding to the A and B positions along the slit. An overall signal-to-noise ratio of 30–50 per resolution element on the final spectra was measured.
As an example, Fig.~\ref{spectral_lines} shows the X-shooter spectrum in selected orders of one observed giant star. Some lines of interest for the abundance analysis are marked.

Table~\ref{tab:1} lists the observed target stars, their coordinates, photometric properties, and distances from the center of Liller~1. The J and K magnitudes are taken from our compilation of NIR photometry \citep{valenti10,ferraro_21} and from the VISTA Variables in the Via Lactea (VVV, \citealt{minniti_10}). 

%%%%%%%%%%%%%%%%%%%%%%%%%%%%%%%%%%%%%%%%%%%%%%%%%%%%%%
\begin{figure}
    \centering
    \includegraphics[width=1\columnwidth]{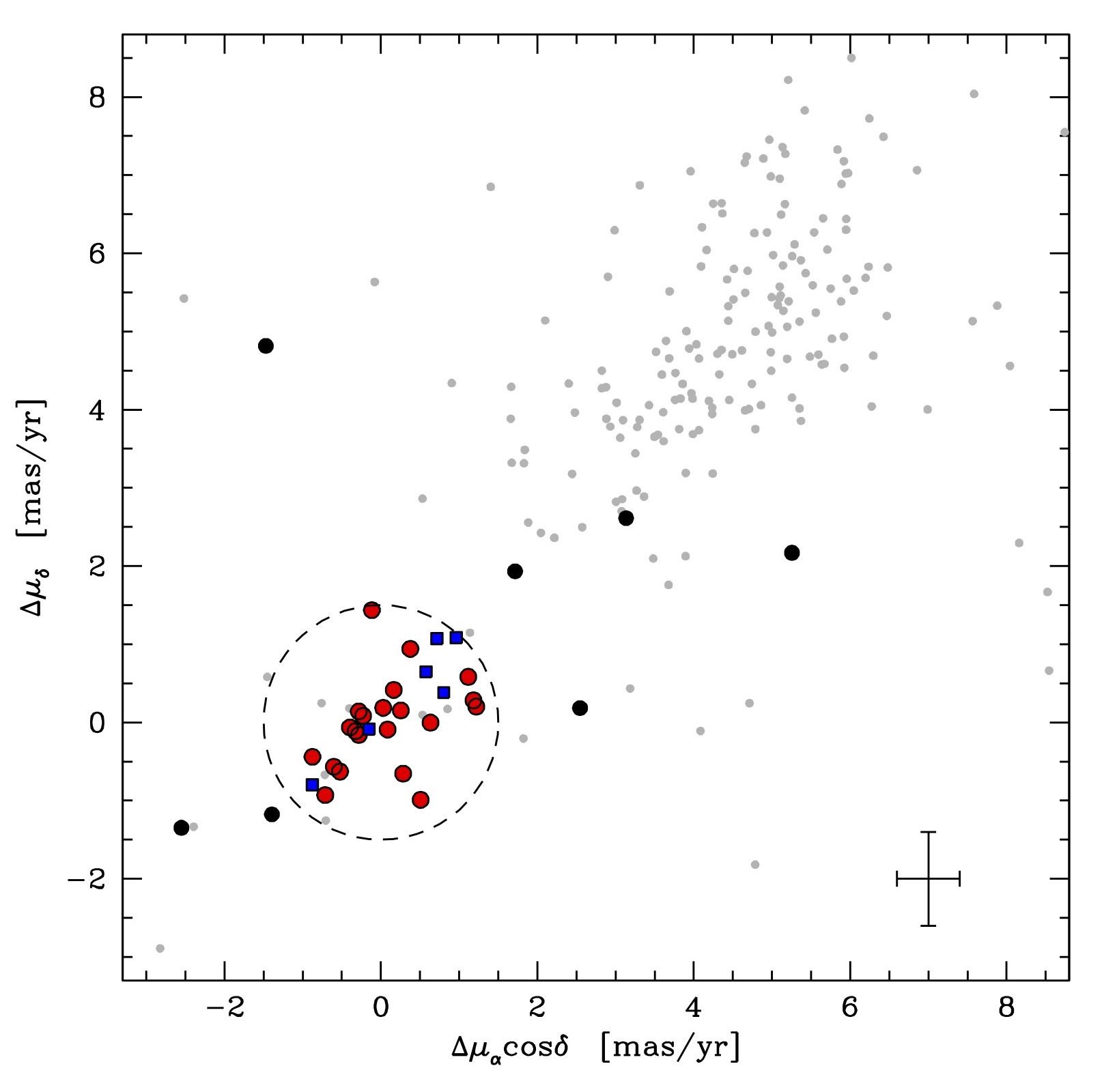}
      \caption{Vector point diagram of the stars with G$<19$ in the direction of Liller~1 (small gray circles), showing the RA and Dec components of the Gaia DR3 proper motions referred to the systemic values quoted by \citet{vasiliev_21}. The large dashed circle is centered on (0,0) and has a radius equal to $3\times\sigma_{\rm PM}$, with $\sigma_{\rm PM}=0.5$ mas yr$^{-1}$ being the proper motion dispersion of Liller~1 member stars. The spectroscopic targets are plotted with large symbols: Large black dots show those classified as Galactic field interlopers due to their discordant proper motions, and large red dots and blue squares show the likely metal-poor and metal-rich members, respectively. The typical error bar is reported in the bottom right corner.}
    \label{vpd}
\end{figure}

\section{Stellar membership from proper motions} 
\label{pm}

\begin{figure*}
    \centering
    \includegraphics[width=\textwidth]{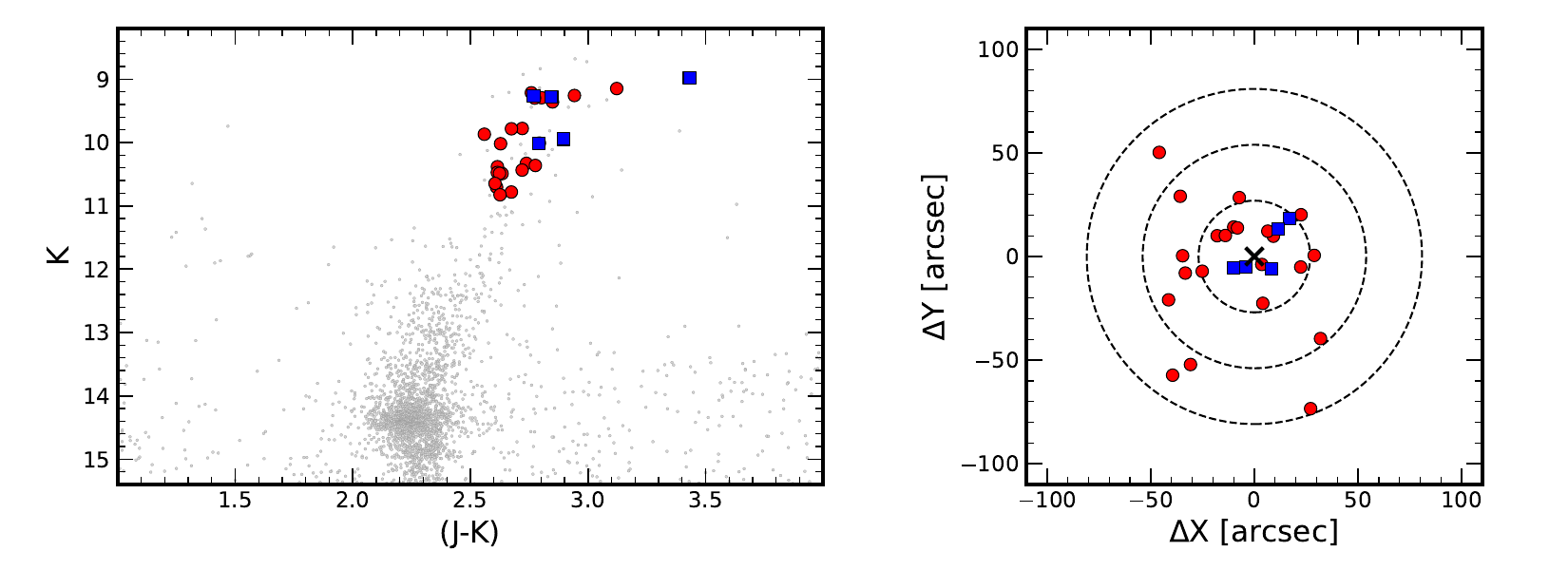}
    \caption{Photometric properties and spatial location of the observed stars. Left panel: K, (J-K) CMD of Liller~1 (gray dots). The likely metal-poor (filled red circles) and metal-rich (filled blue squares) member stars for which we measured chemical abundances (see Sect.~\ref{abundances}) from the observed X-shooter spectra are indicated. 
    Right panel: Distribution of these stars (same symbols) on the plane of the sky with respect to the cluster center, marked with the black cross and located at RA=$263\rlap{.}^\circ3523333$, Dec=$-33\rlap{.}^\circ3895556$. The radii of the dashed black circles are equal to 5, 10, and 15 times the core radius $r_c = 5\rlap{.}^{"}39$ \citep{saracino_15}.
    }
    \label{cmd}
\end{figure*}

The stellar system Liller~1 is located close to the Galactic plane, and field contamination is an issue.  
However, the proper motion distributions plotted in Figs. 2 of \citet{ferraro_21} and \citet{dalessandro_22} show that the Galactic field is clearly distinguishable from the Liller~1 population, which allows a membership selection based on proper motions. The spectroscopic targets are so bright that they are saturated in the HST images that were used in these papers. Hence, we used Gaia DR3 proper motions \citep{gaia_16,gaia_23} in the direction of Liller~1  in order to distinguish member stars from Galactic field interlopers. We found that the proper motion distribution
of the likely member stars in the vector point diagram, centered on the absolute values of the system ($\mu_{\alpha}\text{cos}\delta = -5.403$, $\mu_{\delta} = -7.431$ mas yr$^{-1}$; \citealp{vasiliev_21}), has a dispersion $\sigma_{\rm PM}\approx 0.5$ mas yr$^{-1}$ in both the $\alpha$ and $\delta$ components. 
We therefore assumed that all the observed stars with proper motions within $3\times\sigma_{\rm PM}$ from the absolute systemic values are bona fide members. 
Figure~\ref{vpd} shows the Gaia measurements for the stars with G$<$19 located within 80" from the center of Liller~1(small gray circles). The 34 spectroscopic targets are superposed as large symbols. While the majority (27) of the targets satisfy the adopted membership criterion, being located within $3\times\sigma_{\rm PM}$ in the vector point diagram, 7 stars fall beyond this limit, indicating that they are likely field interlopers. These objects were therefore excluded from the chemical study presented in this paper. Figure~\ref{cmd} shows the location of the 27 target members in the K, J-K CMD (left panel) and on the plane of the sky (right panel) within the central 80" (radius) from the Liller~1 center.

%%%%%%%%%%%%%%%%%%%%%%%%%%%%%%%%%%%%%%%%%%%%%%%%%%%%%%
\section{Spectral analysis} 
\label{analysis}
The X-shooter NIR spectra of the 27 member stars were used to determine radial velocities (RVs) via cross-correlation techniques. The chemical abundances were determined via spectral synthesis.
For this purpose, we used the radiative transfer code TURBOSPECTRUM \citep{alvarez_98,plez_12}, along with MARCS models atmospheres (\citealt{gustafsson_08}), to create grids of synthetic spectra with effective temperatures from 3200 K to 3900 K, log(g)=0.5 dex, $\xi$=2.0 km s$^{-1}$, in a range of metallicities from [Fe/H]$=-1.0$ dex to [Fe/H]$=+0.5$ with a step of 0.1 dex, with two different [$\alpha$/Fe] values, solar scaled ([$\alpha$/Fe]=0.0 dex) and $\alpha$ enhanced ([$\alpha$/Fe]=+0.4 dex). The selected parameters are consistent with those of observed cool giants.
The atomic data were sourced from the VALD3 compilation (\citealt{Ryabchikova_15}), while the most recent molecular data were taken from the website of B. Plez\footnote{\url{https://www.lupm.in2p3.fr/users/plez/}}. 
In order to match the observed line profile broadening, the synthetic spectra were convolved with a Gaussian function at the  R$\approx$8,000 X-shooter resolution. 
This instrumental broadening dominates any other intrinsic broadening, such as macroturbulence and rotation.
For an optimum pixel-to-pixel comparison between the observed and the synthetic spectra, the latter were also resampled to match the pixel size (0.6 $\AA$) of the observed spectra.

\subsection{Stellar parameters}
\label{par}
A photometric estimate of the stellar temperature and gravity was derived from suitable isochrones by \citet{bressan_12} by matching the old (12 Gyr) and young (1,2,3 Gyr) components of Liller~1 at [Fe/H]$\approx$-0.3 and [Fe/H]$\approx$+0.3, respectively, in the observed CMD of \citep{ferraro_21} corrected for differential reddening and assuming their distance modulus (m-M)$_0$=14.65 and average E(B-V)=4.52 as reference.
Low temperatures in the 3400-3800 K range and  gravities in the 0.2-0.8 dex range were obtained. These photometric estimates of the stellar parameters were then cross-checked against the observed OH and CO molecular lines and bandheads in order to simultaneously best fit all of them. 
For the measured member stars, the final adopted temperatures with an uncertainty of $\pm$100~K are listed in Table~\ref{tab:2}, while log~g=0.5$\pm$0.3 dex and a microturbulence value of 2$\pm$0.2 km~s$^{-1}$, typical of luminous bulge giant stars with similar metallicities, were assumed for all of them.
At the low temperatures and gravities of the analyzed stars, the quoted uncertainties for the adopted stellar parameters have an overall impact on the derived abundances of approximately 0.10-0.15 dex. However, it is worth mentioning that this global uncertainty can be regarded as mostly systematic, and it therefore almost canceled out when we computed abundance ratios and when we considered the abundance differences among the Liller~1 stars.
For each star in our sample, we thus generated multiple grids of synthetic spectra with fixed stellar parameters (appropriate to each star) and a varying metallicity from $-1.0$ dex to $+0.5$ dex, in steps of 0.1 dex, with both solar-scaled and some enhancement of [$\alpha$/Fe] and [N/Fe] and corresponding depletion of [C/Fe] for a proper computation of the molecular equilibria, and solar-scaled [X/Fe] values for the other elements.

\subsection{Continuum normalization}
\label{norm}
An optimum continuum normalization of the observed spectra is crucial for determining reliable chemical abundances. 
Throughout the whole grid of synthetic spectra, we therefore considered a local reference continuum by selecting a few wavelength points on the left and right sides of each absorption line of interest that exhibited negligible variation with metallicity. The same wavelength points were selected on the observed spectra. 
We then computed the average "left" and "right" wavelengths and fluxes in the synthetic and observed spectra and performed a linear fit to these "master" points, thus obtaining the best-guess synthetic and observed local continuum.  

%%%%%%%%%%%%%%%%%%%%%
\begin{table*}[t]
\caption{Temperatures, RVs and chemical abundances for the observed stars in Liller~1.}
\label{tab:2}
%\centering
%\resizebox{15cm}{!}{
\scriptsize
\setlength{\tabcolsep}{2.7pt}
\renewcommand{\arraystretch}{1.2}
%\rotatebox{90}{
\begin{tabular}{|c|c|c|c|c|c|c|c|c|c|c|c|c|c|c|}
\hline\hline
ID & T$_{\text{eff}}$ & RV & [Fe/H] & [C/H] & [O/H] & [Na/H] & [Mg/H] & [Al/H] & [Si/H] & [K/H] & [Ca/H] & [Ti/H] & [V/H] & $^{12}$C/$^{13}$C \\
\hline
& [K] & [km s$^{-1}$] & 7.50 & 8.56 & 8.77 & 6.29 & 7.55 & 6.43 & 7.59 & 5.14 & 6.37 & 4.94 & 3.89 & 89 \\
\hline
20  &  3400  &  52  &  -0.28$\pm$0.08  &  -0.42$\pm$0.10  &  +0.19$\pm$0.06  &  +0.32$\pm$0.07  &  +0.12$\pm$0.10  &  +0.08$\pm$0.10  &  +0.02$\pm$0.08  &  +0.14$\pm$0.10  &  +0.21$\pm$0.10  &  +0.26$\pm$0.12  &  -0.05$\pm$0.10  &   5.3$\pm$1.4	  \\
24  &  3400  &  86  &  +0.24$\pm$0.07  &  -0.10$\pm$0.03  &  +0.28$\pm$0.02  &  +0.80$\pm$0.10  &  +0.27$\pm$0.06  &  +0.41$\pm$0.10  &  +0.20$\pm$0.04  &  +0.21$\pm$0.06  &  +0.25$\pm$0.06  &  +0.33$\pm$0.04  &  +0.34$\pm$0.10  &   7.4$\pm$1.7	  \\
27  &  3400  &  69  &  -0.38$\pm$0.06  &  -0.78$\pm$0.04  &  -0.01$\pm$0.10  &  +0.03$\pm$0.03  &  +0.16$\pm$0.05  &  +0.01$\pm$0.05  &  -0.17$\pm$0.04  &  +0.02$\pm$0.05  &  +0.08$\pm$0.10  &  +0.23$\pm$0.08  &  -0.33$\pm$0.10  &   9.2$\pm$1.2	  \\
31  &  3400  &  60  &  -0.24$\pm$0.06  &  -0.74$\pm$0.10  &  +0.09$\pm$0.06  &  +0.17$\pm$0.04  &  +0.06$\pm$0.09  &  +0.03$\pm$0.06  &  +0.07$\pm$0.10  &  +0.11$\pm$0.12  &  +0.03$\pm$0.10  &  +0.31$\pm$0.05  &  -0.21$\pm$0.10  &   9.4$\pm$2.0	  \\
34  &  3400  &  76  &  -0.28$\pm$0.07  &  -1.04$\pm$0.07  &  +0.11$\pm$0.06  &  +0.29$\pm$0.11  &  +0.23$\pm$0.06  &  -0.01$\pm$0.10  &  -0.03$\pm$0.01  &  +0.25$\pm$0.10  &  +0.23$\pm$0.10  &  +0.13$\pm$0.07  &  -0.19$\pm$0.10  &   7.9$\pm$1.3	  \\
35  &  3400  &  86  &  +0.20$\pm$0.06  &  -0.17$\pm$0.03  &  +0.19$\pm$0.06  &  +0.74$\pm$0.10  &  +0.15$\pm$0.10  &  +0.34$\pm$0.10  &  +0.17$\pm$0.06  &  +0.07$\pm$0.10  &  +0.16$\pm$0.10  &  +0.47$\pm$0.05  &  +0.30$\pm$0.10  &  12.3$\pm$1.8	  \\
37  &  3400  &  61  &  -0.33$\pm$0.07  &  -0.58$\pm$0.06  &  +0.11$\pm$0.10  &  -0.09$\pm$0.04  &  +0.18$\pm$0.05  &  +0.07$\pm$0.10  &  -0.14$\pm$0.05  &  -0.06$\pm$0.03  &  +0.13$\pm$0.10  &  +0.13$\pm$0.09  &  -0.37$\pm$0.10  &   6.8$\pm$1.6	  \\
39  &  3400  &  55  &  -0.29$\pm$0.09  &  -0.75$\pm$0.07  &  +0.12$\pm$0.10  &  +0.28$\pm$0.10  &  +0.19$\pm$0.09  &  +0.02$\pm$0.05  &  -0.03$\pm$0.06  &	  --	    &  +0.02$\pm$0.10  &  +0.30$\pm$0.07  &  -0.19$\pm$0.10  &  10.5$\pm$1.3	  \\
48  &  3400  &  42  &  +0.13$\pm$0.08  &  -0.22$\pm$0.06  &  +0.14$\pm$0.10  &  +0.74$\pm$0.10  &  +0.25$\pm$0.10  &  +0.25$\pm$0.06  &  +0.15$\pm$0.05  &  +0.15$\pm$0.10  &  +0.15$\pm$0.10  &  +0.22$\pm$0.07  &  +0.06$\pm$0.10  &  11.3$\pm$1.2	  \\
66  &  3500  &  99  &  -0.06$\pm$0.05  &  -0.20$\pm$0.05  &  +0.09$\pm$0.06  &  +0.57$\pm$0.10  &  +0.19$\pm$0.04  &  +0.18$\pm$0.10  &  +0.14$\pm$0.06  &  +0.21$\pm$0.10  &	     --        &	--	  &  +0.17$\pm$0.10  &  10.9$\pm$1.5	  \\
68  &  3500  &  78  &  -0.36$\pm$0.05  &  -0.63$\pm$0.05  &  -0.06$\pm$0.08  &  -0.04$\pm$0.10  &  +0.06$\pm$0.06  &  +0.05$\pm$0.10  &  -0.19$\pm$0.05  &  -0.12$\pm$0.10  &  +0.03$\pm$0.10  &  +0.18$\pm$0.09  &  -0.43$\pm$0.10  &   7.0$\pm$1.6	  \\
71  &  3500  &  53  &  -0.40$\pm$0.08  &  -0.87$\pm$0.04  &  -0.17$\pm$0.07  &  -0.18$\pm$0.10  &  -0.13$\pm$0.05  &  -0.13$\pm$0.06  &  -0.07$\pm$0.04  &  -0.12$\pm$0.10  &  -0.05$\pm$0.10  &  -0.05$\pm$0.05  &  -0.53$\pm$0.10  &   7.6$\pm$1.0	  \\
74  &  3500  &  70  &  +0.30$\pm$0.06  &  -0.13$\pm$0.03  &  +0.25$\pm$0.07  &  +0.95$\pm$0.10  &  +0.32$\pm$0.10  &  +0.38$\pm$0.10  &  +0.21$\pm$0.05  &  +0.11$\pm$0.10  &  +0.31$\pm$0.10  &  +0.39$\pm$0.05  &  +0.22$\pm$0.10  &  10.5$\pm$1.8	  \\
79  &  3600  &  70  &  -0.22$\pm$0.05  &  -0.97$\pm$0.10  &  -0.05$\pm$0.10  &  +0.01$\pm$0.10  &  +0.11$\pm$0.10  &  +0.12$\pm$0.10  &  +0.06$\pm$0.08  &  +0.30$\pm$0.10  &	     --        &  +0.19$\pm$0.08  &  -0.47$\pm$0.10  &   8.4$\pm$1.0	  \\
85  &  3500  &  61  &  +0.22$\pm$0.07  &  -0.11$\pm$0.10  &  +0.12$\pm$0.06  &  +0.84$\pm$0.10  &  +0.25$\pm$0.05  &  +0.27$\pm$0.10  &  +0.18$\pm$0.04  &  +0.10$\pm$0.10  &  +0.24$\pm$0.10  &  +0.30$\pm$0.10  &  +0.14$\pm$0.10  &   9.4$\pm$1.9	  \\
88  &  3600  &  72  &  -0.34$\pm$0.08  &  -0.67$\pm$0.04  &  +0.12$\pm$0.10  &  +0.02$\pm$0.10  &  +0.05$\pm$0.10  &  +0.12$\pm$0.08  &  -0.05$\pm$0.04  &  +0.06$\pm$0.10  &  +0.02$\pm$0.07  &  +0.07$\pm$0.04  &  -0.29$\pm$0.10  &   8.2$\pm$1.7	  \\
100 &  3700  &  68  &  -0.42$\pm$0.08  &  -1.17$\pm$0.07  &  -0.27$\pm$0.08  &  +0.13$\pm$0.10  &  -0.03$\pm$0.10  &  +0.00$\pm$0.08  &  -0.13$\pm$0.05  &  -0.08$\pm$0.06  &  -0.23$\pm$0.10  &  -0.02$\pm$0.08  &  -0.35$\pm$0.10  &   9.2$\pm$1.7	  \\
103 &  3700  &  40  &  -0.38$\pm$0.07  &  -0.62$\pm$0.04  &  +0.03$\pm$0.06  &  -0.04$\pm$0.10  &  -0.01$\pm$0.10  &  +0.16$\pm$0.05  &  -0.12$\pm$0.05  &  +0.13$\pm$0.10  &  -0.05$\pm$0.10  &  +0.06$\pm$0.04  &  -0.35$\pm$0.10  &  11.5$\pm$2.3	  \\
104 &  3700  &  57  &  -0.38$\pm$0.06  &  -0.68$\pm$0.05  &  +0.01$\pm$0.07  &  -0.04$\pm$0.10  &  -0.08$\pm$0.10  &  -0.07$\pm$0.05  &  -0.16$\pm$0.10  &  +0.12$\pm$0.10  &  -0.14$\pm$0.10  &  -0.08$\pm$0.10  &  -0.42$\pm$0.10  &   9.2$\pm$1.5	  \\
115 &  3700  &  57  &  -0.26$\pm$0.09  &  -0.96$\pm$0.10  &  -0.10$\pm$0.10  &        --	&  +0.15$\pm$0.05  &  +0.19$\pm$0.06  &  -0.02$\pm$0.09  &  -0.08$\pm$0.10  &  +0.01$\pm$0.04  &  +0.21$\pm$0.04  &  -0.21$\pm$0.10  &   8.2$\pm$1.3	  \\
120 &  3700  &  68  &  -0.31$\pm$0.07  &  -0.84$\pm$0.10  &  -0.10$\pm$0.05  &  +0.46$\pm$0.10  &  +0.10$\pm$0.10  &  +0.18$\pm$0.10  &  +0.04$\pm$0.05  &  -0.06$\pm$0.10  &  +0.03$\pm$0.05  &  +0.11$\pm$0.06  &  -0.34$\pm$0.10  &  10.5$\pm$1.4	  \\
121 &  3700  &  62  &  -0.30$\pm$0.07  &  -0.65$\pm$0.04  &  +0.04$\pm$0.06  &  +0.09$\pm$0.10  &  +0.03$\pm$0.10  &  +0.03$\pm$0.10  &  -0.06$\pm$0.04  &  -0.13$\pm$0.10  &  -0.05$\pm$0.05  &  +0.15$\pm$0.04  &  -0.34$\pm$0.10  &  10.9$\pm$2.0	  \\
124 &  3700  &  57  &  -0.32$\pm$0.07  &  -0.67$\pm$0.06  &  +0.04$\pm$0.07  &  +0.05$\pm$0.10  &  +0.07$\pm$0.10  &  +0.02$\pm$0.05  &  -0.03$\pm$0.06  &  +0.14$\pm$0.10  &  +0.13$\pm$0.10  &  +0.19$\pm$0.05  &  -0.31$\pm$0.10  &  10.5$\pm$1.5	  \\
126 &  3700  &  72  &  -0.42$\pm$0.05  &  -0.80$\pm$0.05  &  -0.13$\pm$0.08  &  -0.15$\pm$0.10  &  -0.15$\pm$0.05  &  -0.19$\pm$0.10  &  -0.18$\pm$0.10  &  +0.03$\pm$0.06  &  +0.08$\pm$0.10  &  +0.09$\pm$0.10  &  -0.41$\pm$0.10  &   8.4$\pm$1.1	  \\
132 &  3700  &  66  &  -0.26$\pm$0.08  &  -0.70$\pm$0.05  &  -0.03$\pm$0.07  &  +0.00$\pm$0.10  &  +0.13$\pm$0.05  &  +0.11$\pm$0.06  &  +0.06$\pm$0.07  &  +0.10$\pm$0.06  &  +0.11$\pm$0.06  &  +0.18$\pm$0.05  &  -0.37$\pm$0.10  &   7.6$\pm$1.8	  \\
136 &  3800  &  55  &  -0.29$\pm$0.09  &  -1.06$\pm$0.04  &  -0.10$\pm$0.10  &  +0.37$\pm$0.10  &  +0.10$\pm$0.10  &  +0.07$\pm$0.10  &  -0.10$\pm$0.10  &  +0.16$\pm$0.10  &  +0.03$\pm$0.10  &  +0.16$\pm$0.07  &  -0.33$\pm$0.10  &  10.0$\pm$1.5	  \\
145 &  3800  &  74  &  -0.36$\pm$0.07  &  -0.71$\pm$0.07  &  -0.17$\pm$0.06  &  -0.05$\pm$0.10  &  +0.11$\pm$0.05  &  +0.25$\pm$0.05  &  -0.11$\pm$0.06  &  +0.13$\pm$0.10  &  +0.06$\pm$0.10  &  +0.08$\pm$0.10  &  -0.42$\pm$0.10  &  12.1$\pm$3.1	  \\
\hline\hline
\end{tabular}

\vspace{0.15cm}
Note: The adopted solar abundances for the measured chemical elements are from \citet{magg_22} and they are reported in the header of each element abundance column.
\end{table*}
%\newpage
%%%%%%%%%%%%%%%%%%%%%%%
The observed spectra were then normalized locally through the following formula:
\begin{equation}
    F_{\text{norm}}^{\text{obs}} = F^{\text{obs}}\times \frac{\text{synthetic continuum fit}}{\text{observed continuum fit}},
\end{equation}
where $F^{\text{obs}}$ and $F_{\text{norm}}^{\text{obs}}$ are the observed fluxes before and after the normalization, respectively.
The typical uncertainty in the continuum normalization around each line of interest is $\approx$2\%, and it mostly arises from the photon noise of the few resolution elements that were used to locate it around each line and from the uncertainty in the adopted temperature.

\subsection{Chemical analysis}
\label{chem}
For the chemical analysis, we compiled a list of suitable atomic lines over the entire J, H, K spectral range covered by the X-shooter spectra. This list included lines of Na I, Mg I, Al I, Si I, K I, Ca I, Ti I, V I, and Fe I. We were also able to use OH molecular lines and CO bandheads in order to derive O and C abundances, respectively.  Each line was also scrutinized visually in order to minimize any greater risk of a potential blend with nearby stellar or telluric lines.

We employed a varying number of lines, depending on the chemical element, the signal-to-noise ratio of the spectrum, and possible telluric contamination. For iron, between 4 to 9 lines were analyzed, while for Na, Al, Si, Ca, Ti, and V, between 1 and 5 lines were used. A few OH molecular lines and 2-4 bandheads of CO for the determination of the oxygen and carbon abundance, respectively, were used.

We derived chemical abundances by comparing the synthetic and observed spectra, the latter being optimally normalized around each line of interest, as discussed in Sect.~\ref{norm}. The abundance [X/H] of a given element from each line of interest was determined as the abundance of the synthetic spectrum that minimizes the difference with the observed spectrum. 
As figure of merit, we used the flux of the deepest pixel of each line of interest. However, we also explored solutions that used the sum of the fluxes of the three deepest pixels of each line of interest as a figure of merit, finding abundances that were fully consistent with those obtained using the central pixel alone. Hence, we ultimately decided to rely on the deepest pixel alone to minimize possible residual contamination. 
Random errors in the inferred chemical abundances from each line of interest in the X-shooter spectra are mostly due to the uncertainties in the placement of the continuum (2\%; see Sect.~\ref{norm}) and to the photon noise (2-3\%, according to the measured signal-to-noise ratios of 30-50; see Sect.~\ref{obs}). The overall impact of a random error like this on the derived abundance from each line typically is 0.1-0.2 dex, and this is comparable with the typical 1$\sigma$ scatter in the derived abundances from different lines.
The final errors in the derived abundances, as quoted in Table~\ref{tab:2}, were estimated as the dispersion around the mean abundance value divided by the squared root of the number of lines used to measure each chemical element. When only one line was available, we assumed an error of 0.1 dex, which is the typical line-to-line abundance variation. 
%%%%%%%%%%%%%%%%%%%%%%%%%%%%%%%%%%%%%%%%%%%%%%%%%%%%%%
\begin{figure*}
\centering
\includegraphics[width=0.94\textwidth]{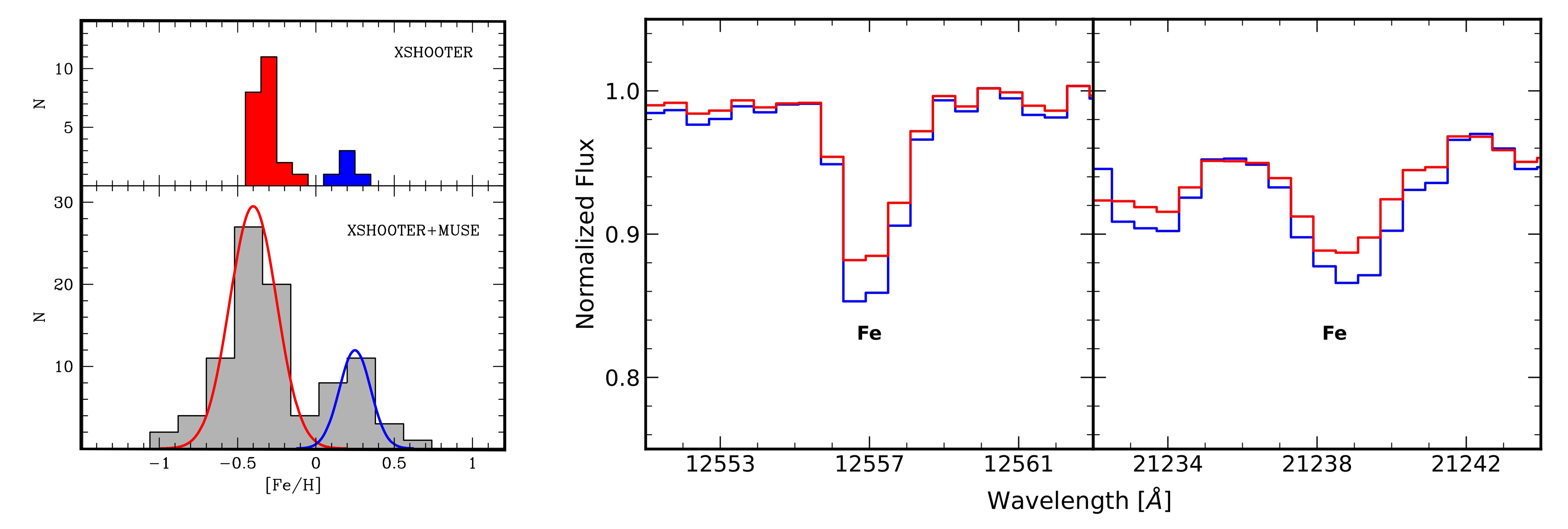}
    \caption{Histograms of the metallicity distribution of Liller~1 (left panel) for the 27 stars observed with X-shooter (top) and for the cumulative sample (from this work and the one by \citealt[][bottom]{Crociati_23}). For the sake of illustration, two Gaussian functions reproducing the subsolar and supersolar components are shown as red and blue curves, respectively. For sake of illustration, X-shooter spectra around two iron lines in the J and K bands for a metal-poor (number 39, red line) and a  metal-rich (number 35, blue line) star of Liller~1, with similar stellar parameters, are also plotted (right panel).}
   \label{distrib_fe}
\end{figure*}

\section{Results}
\label{res}
The analysis of the X-shooter spectra  has provided RVs
with an uncertainty of $<$1~km~s$^{-1}$, chemical abundances [X/H] for eleven elements (Fe, C, O, Na, Mg, Al, Si, K, Ca, Ti, and V)
with typical errors of $\le$0.1~dex, and the $^{12}$C/$^{13}$C isotopic ratios that are listed in Table~\ref{tab:2}. Solar reference abundances are taken from \citet{magg_22}. 
\subsection{Radial velocities}
\label{RV}
The inferred heliocentric RVs  range between 40 and 100 km~s$^{-1}$, with an average value of 65.4$\pm$2.5 km~s$^{-1}$ and a dispersion of 13.2$\pm$1.8 km~s$^{-1}$. These values are fully consistent with the systemic velocity of 67.9$\pm$0.8 km~s$^{-1}$ quoted by \citet{Crociati_23} and within 2.5 times the velocity dispersion of $\approx$13~km~s$^{-1}$ at about 100" from the center\footnote{{\it Fundamental parameters of Galactic globular clusters}, \url{https://people.smp.uq.edu.au/HolgerBaumgardt/globular/}}. 

We therefore conclude that all the 27 proper-motion-selected stars listed in Table~\ref{tab:2} are likely members of Liller~1 according to their 3D kinematics. 

\subsection{Abundances and abundance ratios} \label{abundances}
The distribution of the inferred [Fe/H] values for 27 stars, likely members of Liller~1, is reported in Fig.~\ref{distrib_fe} (left panel). The distribution is clearly bimodal, with a main relatively metal-poor component at an average [Fe/H]=-0.31$\pm$0.02 and 1$\sigma$ dispersion of 0.08$\pm$0.01, including 22 stars, and a metal-rich component at an average [Fe/H]=+0.22$\pm$0.03 and 1$\sigma$=0.06$\pm$0.02, comprising 5 stars.
Interestingly, this measured bimodal distribution agrees with the prediction of \citet{dalessandro_22}.
The stars belonging to the two components are highlighted with different colors in Fig.~\ref{cmd}. 
Figure~\ref{distrib_fe} (right panel) shows the X-shooter spectra around two iron lines in the J and K bands  of a metal-poor and a metal-rich star with similar stellar parameters as an example. The metal-poor star exhibits noticeably shallower features than the metal-rich star, as expected.
The inferred iron abundance dispersion of each subpopulation is consistent with the measurement errors (see Table~\ref{tab:2}).

\begin{figure*}[!h]
\centering
\includegraphics[width=0.94\textwidth]{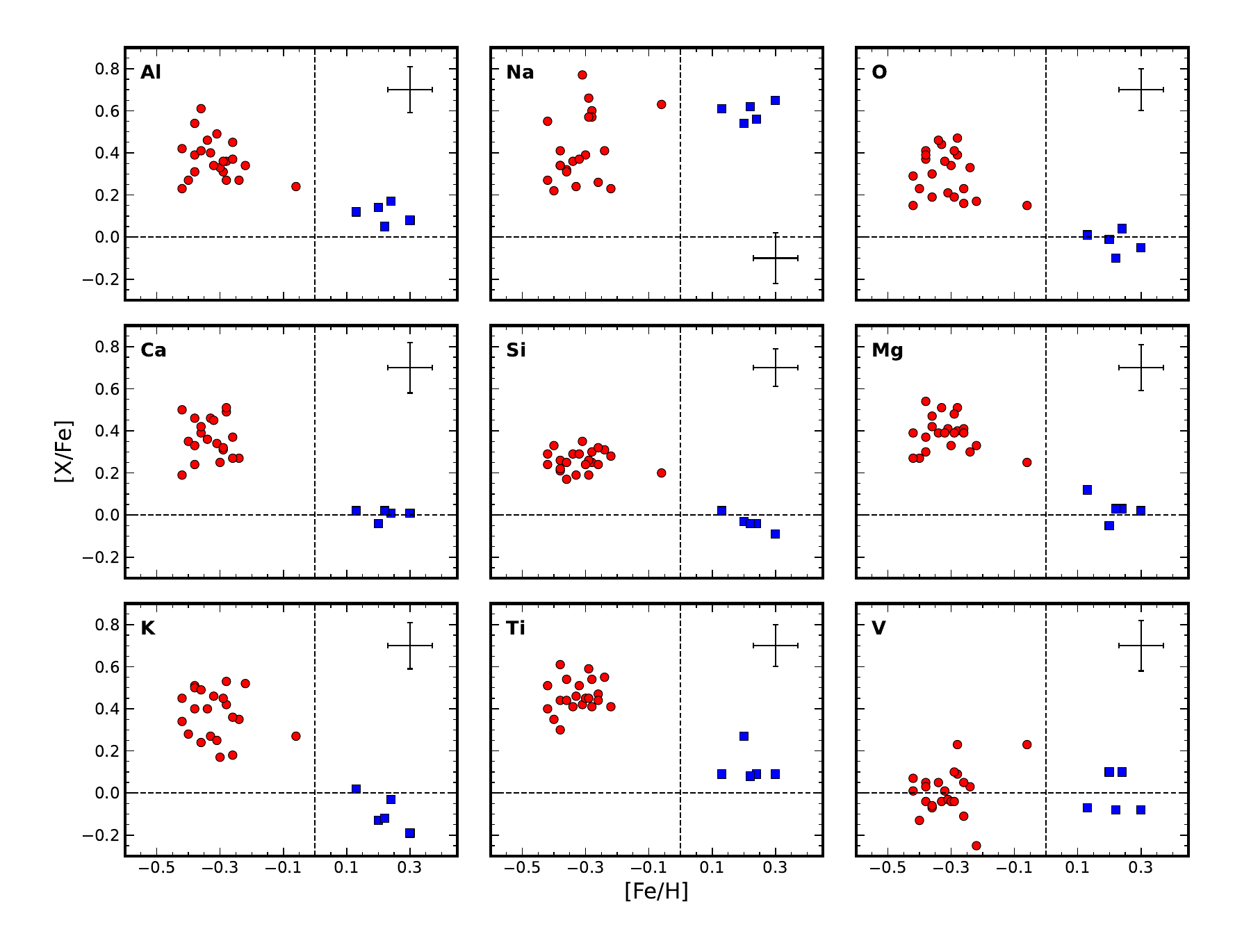}
    \caption{Behavior of [Al/Fe], [Na/Fe], [O/Fe], [Ca/Fe], [Si/Fe], [Mg/Fe], [K/Fe], [Ti/Fe], and [V/Fe] as a function of [Fe/H] for the metal-poor (filled red circles) and metal-rich (filled blue squares) subcomponents we analyzed. The typical error bars of the measurements are reported in the right corner of each panel. The dashed vertical and horizontal lines denote the corresponding zero values.}
   \label{multipanel}
\end{figure*}

\begin{figure*}[!h]
\centering
\includegraphics[width=\textwidth]{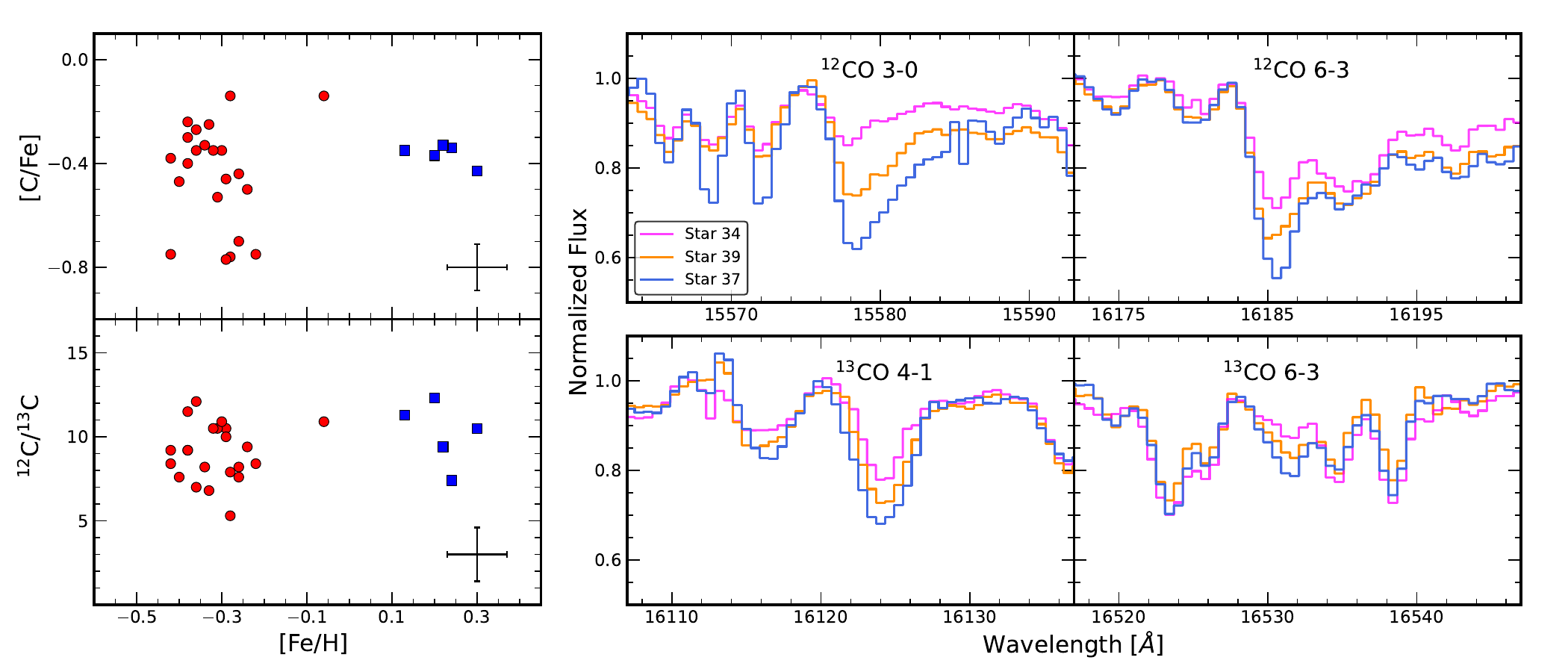}
    \caption{Carbon measurements for the observed stars. Left panel: Behavior of [C/Fe] (top) and $^{12}\textrm{C}/^{13}\textrm{C}$ ratio (bottom) as a function of [Fe/H] for the metal-poor (filled red circles) and metal-rich (filled blue squares) components of Liller~1. The error bars in the bottom right corners represent the typical error associated with the measurements. Right panel: $^{12}$CO 3-0 and 6-3 (top) and $^{13}$CO 4-1 and 6-3 (bottom) roto-vibration molecular bandheads in the H band, as observed in three metal-poor stars with similar stellar parameters, iron abundances, and $^{12}$C/$^{13}$C isotopic ratios, but different carbon abundances.}
   \label{carbon}
\end{figure*}

The  [X/Fe] abundance ratios as a function of [Fe/H] of the other chemical elements measured in the X-shooter spectra are shown in Fig.~\ref{multipanel}. 
The metal-poor subpopulation shows some enhanced (by a factor of 2-3 on average) [$\alpha$/Fe], [Al/Fe], and [K/Fe] with respect to the solar values, while the metal-rich subpopulation shows roughly solar-scaled ratios. The dispersions in these [X/Fe] abundance ratios are consistent with the measurement errors.
The [V/Fe] abundance ratio is about solar at all metallicities, with an average value of 0.01$\pm$0.01 dex and a dispersion of 0.11$\pm$0.01 dex, which is consistent with the measurement errors.
[Na/Fe] is enhanced in the metal-poor and metal-rich subpopulations, 
with average values of 0.42$\pm$0.04 dex and of 0.60$\pm$0.02 dex, respectively, which is consistent with measurements and model predictions for the bulge \citep[see e.g.,][]{johnson_14,kobayashi06,kobayashi11} of some increase in the Na production at high metallicity.
The metal-poor subpopulation shows a dispersion in [Na/Fe] of 0.16$\pm$0.02 dex that only marginally (by $\approx$10\%) exceeds the measurement errors ($\le$0.12 dex), 
while the metal-rich subpopulation has a small dispersion of 0.05$\pm$0.01 dex.

[C/H] abundances and $^{12}$C/$^{13}$C isotopic ratios for all the observed stars (see Table~\ref{tab:2}) were determined by using the unsaturated $^{12}$CO and $^{13}$CO molecular bandheads in the H band. These bandheads are as effective as the individual lines from single roto-vibrational transitions (the latter being only barely if at all distinguishable at the X-shooter resolution) for measuring carbon abundances \citep[see e.g.,][]{fanelli21}.
Figure~\ref{carbon} (left panels) shows the behavior of the [C/Fe] and $^{12}$C/$^{13}$C ratios as a function of [Fe/H].
The metal-poor (red points) and metal-rich (blue points) components of Liller~1 are both significantly depleted in [C/Fe] and in the $^{12}$C/$^{13}$C isotopic ratio (with values in the range of 5-13) with respect to the solar values, 
as expected because of the  mixing and extra-mixing processes in the stellar interiors during the evolution along the RGB. 
The majority of the measured stars in Liller~1 show [C/Fe] depletion well within a factor of four and fully consistent with the values measured in other luminous giants of the bulge and Terzan~5 \citep[see e.g.,][and references therein]{rich12,Origlia_11}. Notably, however, five stars belonging to the metal-poor subpopulations of Liller~1 show an even higher [C/Fe] depletion, up to a factor of seven, while their $^{12}$C/$^{13}$C isotopic ratios (between 8 and 10) are well within the range of values measured in the other Liller~1 giant stars. As an example and for the sake of clarity, Fig.~\ref{carbon} (right panels) shows two $^{12}$CO and two $^{13}$CO molecular bandheads that were used for measuring carbon abundances in three metal-poor stars with similar stellar parameters, iron abundances, and $^{12}$C/$^{13}$C. The $^{12}$CO and the $^{13}$CO bandheads clearly show different depths, according to the different carbon abundances of the three stars. 

\begin{figure}[hb!]
    \centering
    \includegraphics[width=0.99\columnwidth]{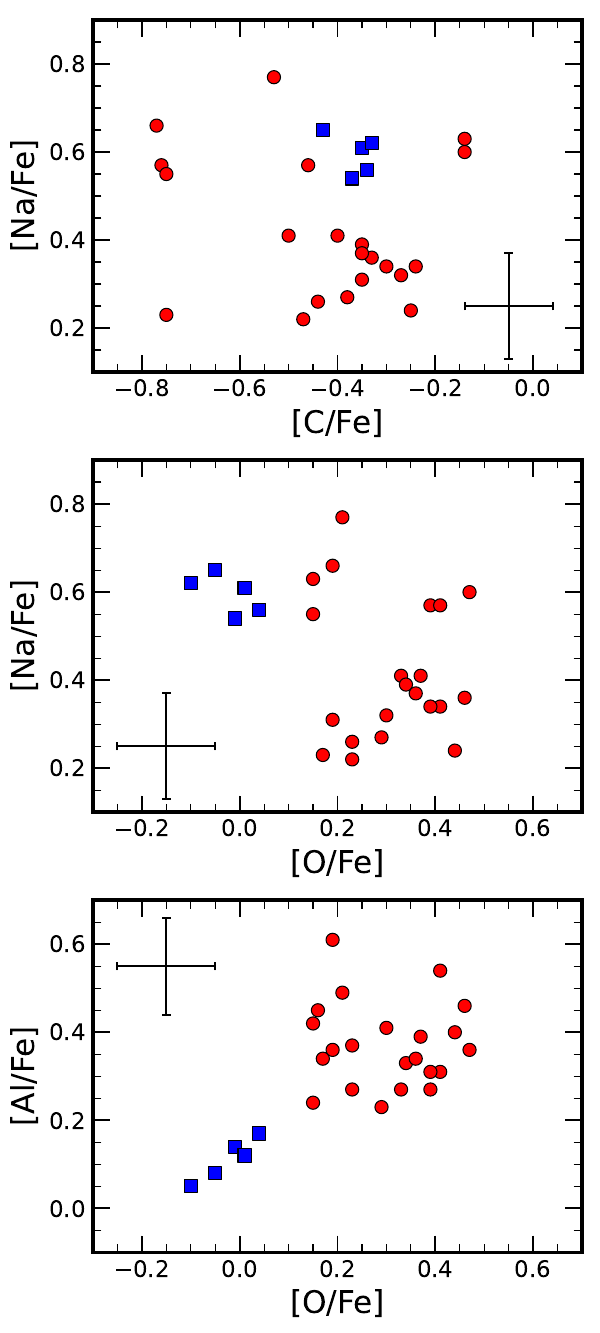}
    \caption{Behavior of [Na/Fe] vs [C/Fe] (top panel) and vs [O/Fe] (middle panel), and of [Al/Fe] vs [O/Fe] for the metal-poor (filled red circles) and metal-rich (filled blue squares) member stars of Liller 1. The typical measurement errors are reported in one corner of each panel.}
    \label{corr}
\end{figure}

Figure~\ref{corr} shows the behavior of  [Al/Fe] (bottom panel) and [Na/Fe] (middle panel) as a function of [O/Fe] and of [Na/Fe] as a function of [C/Fe] (top panel) for the two subpopulations of Liller~1. No specific trend (e.g., an anticorrelation) is evident in the distributions of  these abundance ratios within each subpopulation. The lack of trends within the old subpopulation of Liller~1 in particular is consistent with what is measured in the bulge field and in Terzan~5 \citep{Origlia_11}, and it is at variance with what is typically measured in other old stellar systems, such as genuine globulars (e.g., \citealt{caretta_09}), or in accreted stellar systems such as $\omega$ Centauri \citep{johnson_10,marino_12}, where significant spreads and some anticorrelations among light elements have been found. This evidence therefore has crucial implications for the formation scenario of Liller~1.

\section{Discussion and conclusions}
\label{disc}
Our study of Liller~1 provides the first comprehensive chemical characterization of its two distinct subpopulations: the metal-poor component with subsolar metallicity and enhanced [$\alpha$/Fe], [K/Fe], and [Al/Fe] with respect to solar ratios, and the metal-rich component with supersolar metallicity and about solar scaled ratios for the same light elements.
These chemical properties are consistent with an old age for the metal-poor subpopulation, which likely formed early and quickly from gas enriched by type~II supernovae, and a younger age for the metal-rich subpopulation, which formed from gas that was also enriched by the ejecta of type~Ia supernovae on longer timescales.
Interestingly, according to Fig.~\ref{cmd}, \citep[see also][]{dalessandro_22}, the metal-rich subpopulation is also more centrally concentrated.

Liller~1 is the second complex stellar system in the bulge that, similarly to Terzan~5 \citep[see][]{Ferraro_09, ferraro_16, Origlia_11, Origlia_13, origlia_19, Massari_14}, hosts subpopulations with different ages and metallicities, as well as different [$\alpha$/Fe] abundance ratios. 
Both subpopulations have solar-scaled [V/Fe], thus probing the different metallicities with an additional iron-peak element, and a significantly depleted [C/Fe] and $^{12}$C/$^{13}$C isotopic ratio with respect to the solar values, consistent with mixing and extra-mixing processes during the RGB evolution.
They also show enhanced [Na/Fe], which is consistent with a likely formation and evolution within the bulge.  
Intriguingly, no evidence of the Na-O anticorrelation that is typically observed in globular clusters (GCs; see, e.g., \citealp{caretta_09}) has been found.
This is particular relevant for the subsolar component (shown with red circles in Fig.~\ref{corr}) since it can constrain the formation scenarios of Liller~1. 
This chemical fingerprint has been claimed to be so specific to GCs that it has been proposed as the benchmark to classify a stellar system as a GC \citep{carretta10}. 

Recently, the accretion of a giant molecular cloud by a genuine GC \citep{mckenzie18,bastian22} or the merger of two GCs \citep{khoperskov18,mastrobuono19,pfeffer21} have been proposed as possible scenarios to explain the origin of complex stellar systems such as Terzan~5 and Liller~1. However, in both these scenarios, the subsolar dominant component (tracing the accretor GC) should show the typical GC anticorrelations, which is not the case for Liller~1 (see Fig.~\ref{corr}) nor for Terzan 5 (see  Fig.~3 in \citealt{Origlia_11}). This evidence severely challenges the scenarios that invoke an accretor GC to explain the origin of these systems.
Moreover, events like this are normally very rare, and it is unlikely that they could have occurred multiple times at very specific ages and metallicities, which would be required in order to explain Liller~1 and Terzan~5 multi-iron distributions with at least three main peaks. Even if this were the case, it cannot explain the underlying continuous star formation over almost the entire lifetime of Liller~1.

The suggestion that Terzan~5 and Liller~1 could have an extragalactic origin, that is, that they are the former nuclear star cluster of an accreted dwarf galaxy \citep[see e.g.,][]{brown18,alfaro19,taylor22}, has been almost discarded \citep[see e.g.,][and references therein]{pfeffer21} based on kinematic and age-metallicity considerations. 
Metallicity also makes it very unlikely that Liller~1 and Terzan~5 could have formed in a significantly more metal-poor Galactic environment such as the halo \citep{moreno22}.

All these facts thus suggest that Liller~1 and Terzan~5 are likely complex stellar systems that have formed and evolved within the bulge.
They are both very massive, with a present-day mass exceeding 10$^6$~M$_{\odot}$ \citep{Lanzoni10, ferraro_21}, but they probably were more massive in the past and were thus able to retain the SN ejecta and possibly to self-enrich, as shown by the recent chemical evolution modeling of Terzan~5 presented in \citet{romano23}.
Hence, they could be fossil fragments of the pristine clumps of stars and gas that may have contributed to forming the early bulge \citep[e.g.,][and references therein]{immeli_04, elemgreen_08}.
These fragments could have survived complete disruption and could have evolved and self-enriched as independent systems within the bulge. They might also have been able to experience some new events of star formation at later epochs, as probed by their younger and more metal-rich subpopulations, and  consistent with the reconstructed SFH of Liller~1 \citep{dalessandro_22}.

Given the importance of this complex but intriguing stellar systems in shedding light on the bulge formation and evolution, detailed chemical studies of Terzan~5 and Liller~1 at higher spectral resolution are ongoing. The results will be presented in forthcoming papers.

%%%%%%%%%%%%%%%%%%%%%%%%%%%%%%%%%%%%%%%%%%%%%%%%%%%%%%

\begin{acknowledgements}
CF and LO acknowledge the financial support by INAF within the VLT-MOONS project.    
This work is part of the project Cosmic-Lab at the Physics and Astronomy Department “A. Righi” of the Bologna University (http:// www.cosmic-lab.eu/ Cosmic-Lab/Home.html).
\end{acknowledgements}
\bibliographystyle{aa} % style aa.bst
\bibliography{biblio} % your references Yourfile.bib

\end{document}